\documentclass[preprint2]{aastex}
\newcounter{rom}

\newcommand{\gal}{CH$_2$OHCHO}
\newcommand{\gto}{\object{G31.41+0.31}}

\shorttitle{On the formation of glycolaldehyde}
\shortauthors{Woods et al.}

\begin{document}


\title{On the formation of glycolaldehyde \\
    in dense molecular cores}


\author{Paul M. Woods, George Kelly and Serena Viti}
\affil{Department of Physics \& Astronomy, University College London, Gower Street, London WC1E 6BT, UK}
\email{pmw@star.ucl.ac.uk}

\and

\author{Ben Slater, Wendy A. Brown, Fabrizio Puletti, Daren J. Burke and Zamaan Raza}
\affil{Department of Chemistry, University College London, 20 Gordon Street, London WC1H 0AJ, UK}







\begin{abstract}
Glycolaldehyde is a simple monosaccharide sugar linked to prebiotic
chemistry. Recently it was detected in a molecular core in the
star-forming region \gto\ at a reasonably high abundance. We
investigate the formation of glycolaldehyde at 10\,K to determine
whether it can form efficiently under typical dense core
conditions. Using an astrochemical model, we test five different
reaction mechanisms that have been proposed in the astrophysical
literature, finding that a gas-phase formation route is unlikely. Of
the grain-surface formation routes, only two are efficient enough at
very low temperatures to produce sufficient glycolaldehyde to match
the observational estimates, with the mechanism culminating in
CH$_3$OH + HCO being favoured. However, when we consider the
feasibility of these mechanisms from a reaction chemistry perspective,
the second grain-surface route looks more promising, H$_3$CO + HCO.

\end{abstract}





\section{Introduction}

\begin{deluxetable}{llcc}
\tabletypesize{\scriptsize}
\tablecaption{Summary of proposed reaction pathways \label{tab:schemes}}
\tablewidth{0pt}
\tablehead{
\colhead{Reaction} & \colhead{Reference} & \colhead{Medium} & \colhead{Method} } 
\startdata
{\bf A1.} $g$--H$_2$O + h$\nu$\ $\longrightarrow$\ $g$--OH + $g$--H & & &  \\
{\bf A2.} $g$--CH$_4$\ +  h$\nu$\ $\longrightarrow$\ $g$--CH$_3$\ + $g$--H & & & \\
{\bf A3.} $g$--CH$_3$\ +  $g$--OH $\longrightarrow$\ $g$--CH$_3$OH & \citet{sor01} & grain mantle & theory\\
{\bf A4.} $g$--CO + $g$--H $\longrightarrow$\ $g$--HCO & & (H$_2$O/CH$_4$/NH$_3$/CO) & \\
{\bf A5.} $g$--CH$_3$OH + $g$--HCO $\longrightarrow$\ $g$--\gal\ + $g$--H & & & \\
\tableline
{\bf B1.} $g$--CH$_3$OH + CRP $\longrightarrow$\ $g$--CH$_2$OH + $g$--H & & & \\
{\bf B2.} $g$--CO + $g$--H $\longrightarrow$\ $g$--HCO & \citet{ben07b} & grain mantle & experiment \\
{\bf B3.} $g$--CH$_2$OH + $g$--HCO $\longrightarrow$\ $g$--\gal & & (CH$_3$OH/CO) \\
\tableline
{\bf C1.} H$_3^+$\ + H$_2$CO $\longrightarrow$\ H$_2$COH$^+$\ + H$_2$ & \\
{\bf C2.} H$_2$COH$^+$\ + H$_2$CO $\longrightarrow$\ CH$_2$OHCH$_2$O$^+$ & \citet{hal06} & gas & theory \\
{\bf C3.} CH$_2$OHCH$_2$O$^+$\ $\longrightarrow$\ CH$_2$OHCHOH$^+$ & \\
{\bf C4.} CH$_2$OHCHOH$^+$\ $\longrightarrow$\ \gal\ + H$^+$ & \\
\tableline
{\bf D1.} $g$--CO + $g$--H + $g$--H $\longrightarrow$\ $g$--H$_2$CO & \\
{\bf D2.} $g$--CO + $g$--H $\longrightarrow$\ $g$--HCO & \citet{bel09} & surface & theory \\
{\bf D3.} $g$--H$_2$CO + $g$--HCO + $g$--H $\longrightarrow$\ $g$--\gal & \\
\tableline
{\bf E1.} $g$--CO + $g$--H $\longrightarrow$\ $g$--HCO & \\
{\bf E2.} $g$--HCO + $g$--C $\longrightarrow$\ $g$--HC$_2$O & \\
{\bf E3.} $g$--HC$_2$O + $g$--H $\longrightarrow$\ $g$--CH$_2$CO & \citet{cha05} & surface & theory \\
{\bf E4.} $g$--CH$_2$CO + $g$--H $\longrightarrow$\ $g$--CH$_2$CHO & \\
{\bf E5.} $g$--CH$_2$CHO + $g$--O $\longrightarrow$\ $g$--OCH$_2$CHO & \\
{\bf E6.} $g$--OCH$_2$CO + $g$--H $\longrightarrow$\ $g$--CH$_2$OHCHO & \\
\enddata
\tablecomments{$g$-- signifies a grain-surface species, h$\nu$\
signifies a UV photon and CRP signifies a cosmic ray particle. }
\end{deluxetable}

The chemistry of dense molecular cores -- the birth sites of massive
stars -- is demonstrably complex, in that large molecules composed of
several functional groups are observed to be present. Of particular
interest for their astrobiological implications \citep{rem04,sny06}
are the isomers of composition C$_2$H$_4$O$_2$, viz. methyl formate,
acetic acid, and glycolaldehyde. Glycolaldehyde (\gal), a simple
monosaccharide sugar linked with the formation of RNA and amino acids
in terrestrial environments \citep{col95,web98}, was detected first
towards the Galactic Centre molecular cloud \object{Sagittarius B2(N)}
\citep{hol00}, and more recently towards a star-forming hot molecular
core, \gto\ \citep{bel09}, both rich sources of molecules.

The mechanism of glycolaldehyde formation in these environments is
uncertain, although it is becoming increasingly clear that the
site of the formation of large organic molecules is the icy surfaces
of astronomical dust \citep[e.g.,][]{gar06}. As suggested by early
models of grain-surface chemistry, much of the development of complex
molecules is through fairly rapid hydrogenation of frozen-out
gas-phase molecules \citep[e.g.,][]{tie97}. Once the end-points of
these processes have been reached (e.g., C$\rightarrow$CH$_3$OH,
N$\rightarrow$NH$_3$) molecules and radicals must move through the ice
lattices in order to build the large organic molecules we detect in
regions of star formation.

In this paper we investigate the formation of glycolaldehyde in a
collapsing cloud core at 10\,K by comparing five mechanisms that have
been suggested in the astrophysical literature. These mechanisms are
highly speculative; all are without associated reaction rate
coefficients and many of the reactions involved have not previously
been included in astrochemical models to assess their
effectiveness. We aim to constrain the possible formation routes of
glycolaldehyde in cold cores, by investigating the wide parameter
space resulting from the lack of existing constraints. This work is an
initial investigation which forms part of a larger program looking
into the formation of glycolaldehyde in the dense interstellar medium,
using the combined tools of astrochemical modelling, experimental
surface chemistry and quantum chemical calculations.

Dense prestellar cores have a very limited range of temperatures, from
$\sim$7--11\,K \citep[e.g.,][]{pag07,ber06,lai03,hot02}, and there is
experimental evidence \citep{obe09,ben07b} that glycolaldehyde forms
in such low temperature environments. Given the uncertainties in
reaction rates, we look at a large parameter space, and conservatively
restrict ourselves to simple hydrogenation of the species which are
frozen out onto grain surfaces. In this way we identify which of the
mechanisms suggested in an {\it ad hoc} manner are feasible for
the production of glycolaldehyde in molecular cores such as \gto. In
\S\ref{sec:mechs} we give details on the selected mechanisms which we
investigate. \S\ref{sec:model} gives an overview of our model, and the
procedure which we follow in investigating the mechanisms. In
\S\ref{sec:results} and \S\ref{sec:discuss} we draw out some results
from our modelling, and evaluate them bearing the chemical energetics
of the reactions in mind. In \S\ref{sec:concs} we conclude with a
summary of our findings.

\section{Proposed pathways to glycolaldehyde}
\label{sec:mechs}

Since its detection in space, there has been significant interest in
glycolaldehyde formation \citep[e.g.][ and
  others]{sor01,cha05,hal06,ben07b,bel09}. Several mechanisms have
been proposed, including both gas-phase and surface reactions, and
experiments have been conducted on laboratory surface analogs. We
summarise some of the work which has been carried out below, and in
Table~\ref{tab:schemes}. We do not consider high temperature
($\sim$300\,K) formation routes, e.g., \citet{jal07}.

Some of these reactions have been tested in hot core models, at
temperatures up to 200\,K. For example, \citet{gar08} \citep[and,
  presumably,][]{laa11} incorporate reactions B3 and E6. However, the
chemistry in these warm temperature regimes is somewhat different,
since surface radicals can be sufficiently energetic to overcome
diffusion barriers, affording them greater mobility on grain
surfaces. We only consider these reactions at 10\,K in order to test
whether glycolaldehyde can form efficiently in the isothermal collapse
phase of star formation.

Below we summarise the work from which the reaction mechanisms in
Table~\ref{tab:schemes} comes.

\subsection{Mechanism A}

\citet{sor01} discusses the theory of processing icy grain mantles in
the interstellar medium with ultraviolet (UV) radiation, producing
high concentrations of free radicals (particularly OH and
CH$_3$). These radicals then react in the grain mantles in order to
produce large organic molecules such as amino acids and sugars, with
the energy for thermal hopping coming from grain-grain collisions. The
resulting large organics (including glycolaldehyde) would then be
desorbed into the gas phase following mantle explosions, and despite
some fraction of these large molecules being destroyed in the process,
some would remain intact.

\subsection{Mechanism B}

\citet{ben07b} simulated the bombardment of grain mantles with cosmic
ray particles by irradiating laboratory methanol/carbon monoxide ices
with energetic electrons at 11\,K. Cosmic rays can penetrate entire
grains, producing up to 100 suprathermal particles each, which then
ionise (methane) ice molecules \citep{kai97,kai02}. The resulting
high-energy electrons ($\sim$5\,keV) may then affect the mantle
chemistry by forming radicals, which subsequently react to form large
organic molecules. In a methanol/carbon monoxide ice these large
organics include C$_2$H$_4$O$_2$\ isomers. The experiment showed that
both glycolaldehyde and methyl formate were formed, in addition to
many smaller molecules and radicals. Acetic acid was not detected, but
can be formed in methane/carbon dioxide ices \citep{ben07a}.

\subsection{Mechanism C}

\citet{hal06} postulate that glycolaldehyde may be formed in the gas
phase through acid-catalysed reactions of formaldehyde, based on
research on formose reactions \citep{but61,bre59}.  Formaldehyde would
react with its protonated form to create an intermediate species,
which would then undergo reorganisation into protonated
glycolaldehyde. There is some experimental evidence for this method,
although it is unclear whether the resulting C$_2$H$_4$O$_2$\ isomer
is in fact glycolaldehyde \citep{jal07}.

\subsection{Mechanism D}

\citet{bel09} highlighted the potential importance of the HCO radical
in glycolaldehyde formation. They suggested that reactions between HCO
and methanol (or methanol derivatives) or formaldehyde could occur
rapidly on grain surfaces in hot cores. Gas-phase routes would be too
inefficient. The simplicity of the reaction pathway, which is driven
by rapid hydrogenation and the reaction of small surface radicals,
means that glycolaldehyde formation could be efficient when densities
are high. Only small amounts of CO would need to be processed on
grains.

\subsection{Mechanism E}

\citet{cha05} suggested that complex molecules build up on grain
surfaces through the aggregation of common atoms, since at low
temperatures only atoms are likely to be mobile. At early times, atoms
such as C, N or O may accrete significantly, whereas at late times,
when most heavy atoms will have frozen out, hydrogenation of molecules
will dominate. Such a scheme could not only lead to the formation of
glycolaldehyde, but also to other large molecules such as acetic acid
and aminomethanol. Methyl formate, however, cannot be formed through
this kind of pathway, but only through the combination of relatively
large surface radicals.

\section{The chemical model and scientific procedure}
\label{sec:model}

In order to test which of the suggested routes to glycolaldehyde
formation are feasible in dense core environments, we have
incorporated the above chemical reactions (Table~\ref{tab:schemes})
into a model of a hot molecular core. The model, described below in
more detail, is based on that described in \citet{vit04}.

\subsection{The model}

\begin{deluxetable}{lcc}
\tabletypesize{\footnotesize}
\tablecaption{Hydrogenation percentages for accreting species \label{tab:hydrog}}
\tablewidth{0pt}
\tablehead{
\colhead{Accreting species} & \colhead{Products in regime 1 (f1)} & \colhead{Products in regime 2 (f2)} } 
\startdata
O   & 2\% O, 18\% OH, 80\% H$_2$O & 1\%O, 9\% OH, 90\% H$_2$O \\
CO  & 70\% CO, 20\% HCO, 5\% H$_2$CO, 5\% CH$_3$OH & 60\% CO, 25\% HCO, 10\% H$_2$CO, 5\% CH$_3$OH \\
C   & 2\% C, 3\% CH, 5\% CH$_2$, 20\% CH$_3$, 70\% CH$_4$ & 1\% C, 4\% CH, 8\% CH$_2$, 12\% CH$_3$, 75\% CH$_4$ \\
HCO$^{(+)}$ & 20\% HCO, 40\% H$_2$CO, 40\% CH$_3$OH &  10\% HCO, 45\% H$_2$CO, 45\% CH$_3$OH \\
OH & 10\% OH, 90\% H$_2$O &  5\% OH, 95\% H$_2$O \\
\enddata
\end{deluxetable}

\setcounter{rom}{1} The model is a two-phase time-dependent model
which follows the collapse of a prestellar core (phase
\Roman{rom}),\addtocounter{rom}{1} followed by the subsequent warming
and evaporation of grain mantles (phase
\Roman{rom}).\addtocounter{rom}{-1} We only consider phase
\Roman{rom}, since in this work we wish to investigate the potential
formation of glycolaldehyde at low temperatures, as suggested by
experiment \citep{obe09,ben07b}. In phase \Roman{rom}, a diffuse cloud
of density 10$^2$\,molecules cm$^{-3}$\ undergoes free-fall collapse
until it has reached a density of $\sim$10$^7$\,cm$^{-3}$. This occurs
on a timescale of half a million years and at a temperature of
10\,K. During the collapse, atoms and molecules collide with, and
freeze on to, grain surfaces. We assume that hydrogenation occurs
rapidly on these surfaces, so that, for example, some percentage of
carbon atoms accreting will rapidly become frozen-out methane,
CH$_4$. Initial atomic abundances are taken from \citet{sof01}, as in
\citet{vit04}. We employ the reaction rate data from the
UDfA06\footnote{\url{http://www.udfa.net}} astrochemical database,
augmenting it with grain-surface (hydrogenation) reactions and those
reactions included in Table~\ref{tab:schemes}. In the formation of
glycolaldehyde we only consider the most propitious of
circumstances. We discount the destruction of glycolaldehyde on the
grains through cosmic ray strikes, photodissociation or further
reaction, so that the quantity of glycolaldehyde formed can be
regarded as an upper limit. We also assume that due to the cold
temperature, no species will desorb from the grains except H and
He. This assumption is reasonable when considering thermal desorption,
but neglects the effect of non-thermal desorption mechanisms on both
glycolaldehyde and the reactants which go into its formation. Any
proposed reaction pathways that do not produce reasonable amounts of
glycolaldehyde under these conditions can surely be dismissed from
consideration.

\subsection{The procedure}

\begin{deluxetable}{lll}
\tabletypesize{\footnotesize}
\tablecaption{Adopted reaction rates for key reactions \label{tab:rates}}
\tablewidth{0pt}
\tablehead{
\colhead{Reaction} & \colhead{$\alpha$ coefficient} & \colhead{Comment} } 
\startdata
{\bf A1.} $g$--H$_2$O + h$\nu$\ $\longrightarrow$\ $g$--OH + $g$--H & 5.9$\times$10$^{-10}$ ($\gamma$=1.7) & Based on gas-phase\\
{\bf A2.} $g$--CH$_4$\ +  h$\nu$\ $\longrightarrow$\ $g$--CH$_3$\ + $g$--H & 2.2$\times$10$^{-10}$ ($\gamma$=2.2)  & Based on gas-phase \\
{\bf A3.} $g$--CH$_3$\ +  $g$--OH $\longrightarrow$\ $g$--CH$_3$OH & 1.0$\times$10$^{-9}$ & Based on similar gas-phase\\
{\bf A4.} $g$--CO + $g$--H $\longrightarrow$\ $g$--HCO & \ldots & Assumed fast \\
{\bf A5.} $g$--CH$_3$OH + $g$--HCO $\longrightarrow$\ $g$--\gal\ + $g$--H & 3.0$\times$10$^{-17}$ & Methyl formate rate retarded by $\times$50\\
\tableline
{\bf B1.} $g$--CH$_3$OH + CRP $\longrightarrow$\ $g$--CH$_2$OH + $g$--H & 1.3$\times$10$^{-17}$ & Standard $\zeta$\ (CRP = cosmic ray)  \\
{\bf B2.} $g$--CO + $g$--H $\longrightarrow$\ $g$--HCO & \ldots & Assumed fast \\
{\bf B3.} $g$--CH$_2$OH + $g$--HCO $\longrightarrow$\ $g$--\gal &  3.0$\times$10$^{-17}$  & Conservative estimate\\
\tableline
{\bf C1.} H$_3^+$\ + H$_2$CO $\longrightarrow$\ H$_2$COH$^+$\ + H$_2$ & 6.3$\times$10$^{-9}$  &  Based on similar gas-phase\\
{\bf C2.} H$_2$COH$^+$\ + H$_2$CO $\longrightarrow$\ CH$_2$OHCH$_2$O$^+$ & 1.0$\times$10$^{-10}$ & Based on similar gas-phase\\
{\bf C3.} CH$_2$OHCH$_2$O$^+$\ $\longrightarrow$\ CH$_2$OHCHOH$^+$ & \ldots & Assumed fast\\
{\bf C4.} CH$_2$OHCHOH$^+$\ $\longrightarrow$\ \gal\ + H$^+$ & \ldots  & Assumed fast \\
\tableline
{\bf D1.} $g$--CO + $g$--H + $g$--H $\longrightarrow$\ $g$--H$_2$CO & \ldots & Assumed fast\\
{\bf D2.} $g$--CO + $g$--H $\longrightarrow$\ $g$--HCO & \ldots & Assumed fast \\
{\bf D3.} $g$--H$_2$CO + $g$--HCO + $g$--H $\longrightarrow$\ $g$--\gal & 3.0$\times$10$^{-17}$  & Conservative estimate, based on\\
 & &  methyl formate reaction\\
\tableline
{\bf E1.} $g$--CO + $g$--H $\longrightarrow$\ $g$--HCO & \ldots & Assumed fast \\
{\bf E2.} $g$--HCO + $g$--C $\longrightarrow$\ $g$--HC$_2$O &  1.0$\times$10$^{-10}$  & Based on similar gas-phase\\
{\bf E3.} $g$--HC$_2$O + $g$--H $\longrightarrow$\ $g$--CH$_2$CO & 1.0$\times$10$^{-9}$ & Based on similar gas-phase \\
{\bf E4.} $g$--CH$_2$CO + $g$--H $\longrightarrow$\ $g$--CH$_2$CHO & 5.0$\times$10$^{-9}$ & Based on similar gas-phase \\
{\bf E5.} $g$--CH$_2$CHO + $g$--O $\longrightarrow$\ $g$--OCH$_2$CHO &  1.0$\times$10$^{-12}$ & Based on H-atom gas phase reaction,\\
& & retarded\\
{\bf E6.} $g$--OCH$_2$CO + $g$--H $\longrightarrow$\ $g$--CH$_2$OHCHO & 3.0$\times$10$^{-17}$  & Conservative estimate, based on\\
& &  methyl formate reaction\\
\enddata
\tablecomments{Reactions A4, B2, D2 and E1 are identical.}
\end{deluxetable}

Firstly, we investigate the five schemes individually, varying key
model parameters, such as the rate coefficients, the final collapse
density, the incident UV field, the cosmic ray ionisation rate and the
hydrogenation efficiency of accreted molecules. Finally we look at all
the mechanisms together, to find which is the most efficient in
competition.

Many of the rate coefficients of the reactions in
Table~\ref{tab:schemes} are completely unknown. In light of this, we
consider a wide parameter space, covering up to 14 orders of magnitude
in reaction rate. Our aim is to understand the behavior of the
reactions in each mechanism, and what effect they have on the
abundance of glycolaldehyde, not to determine reaction rates. However,
through our investigation we may be able to better constrain possible
reaction rates. Where practical, we adopt identical or similar
gas-phase reaction rates from UDfA06 or
KIDA\footnote{\url{http://kida.obs.u-bordeaux1.fr}} for unknown
grain-surface rates as a conservative initial estimate
(Table~\ref{tab:rates}), given that the grain surface is thought to
act as a catalyst. Glycolaldehyde is not included in the standard
UDfA06 database, so we utilise rates from analogous reactions which
produce methyl formate, or we make very conservative estimates. In
varying the rates, we vary only the $\alpha$-parameter in the rate
coefficients: $k = \alpha(T/300\mathrm{K})^\beta\exp(-\gamma/T)$\ for
two-body reactions, $k = \alpha$\ for reactions with cosmic rays, and
$k = \alpha\exp(-\gamma A_\mathrm{V})$\ for photoreactions.

We test two final collapse densities, $n_\mathrm{f}=10^6$\ and
10$^7$\,cm$^{-3}$, which has implications also for the freeze-out
percentage of molecules. \Citet{vit01} argued that in hot cores
freeze-out is never total and in fact some gaseous CO is always
observed, even in regions where no millimetre continuum is detected
\citep[e.g.,][]{mol00}. Hence we impose the constraint that a maximum
of 90\% of the circumstellar material is frozen out at a final density
of $n_\mathrm{f}=10^7$\,cm$^{-3}$, and 75\% at 10$^6$\,cm$^{-3}$.

The strength of the impinging UV field within the core was adjusted
solely for mechanism A, since it involves the UV processing of
molecules on grains. We investigated the effects of scaling the
standard interstellar UV field strength, G$_\mathrm{0}$, by up to 30
times. When investigating reaction B1, we increase the cosmic ray
ionisation rate in the model globally by up to 1\,000 times.

Finally, we tested two different grain-surface hydrogenation regimes,
one where the products were more saturated (\textquotedblleft
f2\textquotedblright), and one less saturated (\textquotedblleft
f1\textquotedblright),. These regimes are represented in
Table~\ref{tab:hydrog}. Hydrogenation rates on grain surfaces are
unknown, but the calculation of hydrogen-atom hopping and tunnelling
rates show that they are rapid in comparison to other grain-surface
reactions \citep{gou07,tie89}. At low temperatures, the recombination
of physisorbed atomic hydrogen with chemisorbed atoms dominates. At
temperatures $<$20\,K, molecular hydrogen formation efficiency on
grain surfaces is near unity \citep{caz04,wil07}. We assume that
hydrogenation of species on grain surfaces is instantaneous.

We combine these free parameters into a grid of model results.

\section{Results}
\label{sec:results}

We have calculated approximately 450 models to investigate the
formation of glycolaldehyde at 10\,K during the isothermal collapse
phase of star formation. Many of these permutations arise from varying
the $\alpha$\ parameter in the rate coefficient of the reactions
involved in the mechanisms by a factor of up to 10$^{\pm7}$.

\subsection{The effect of scaling the UV field strength on mechanism A}

\begin{figure}
\plotone{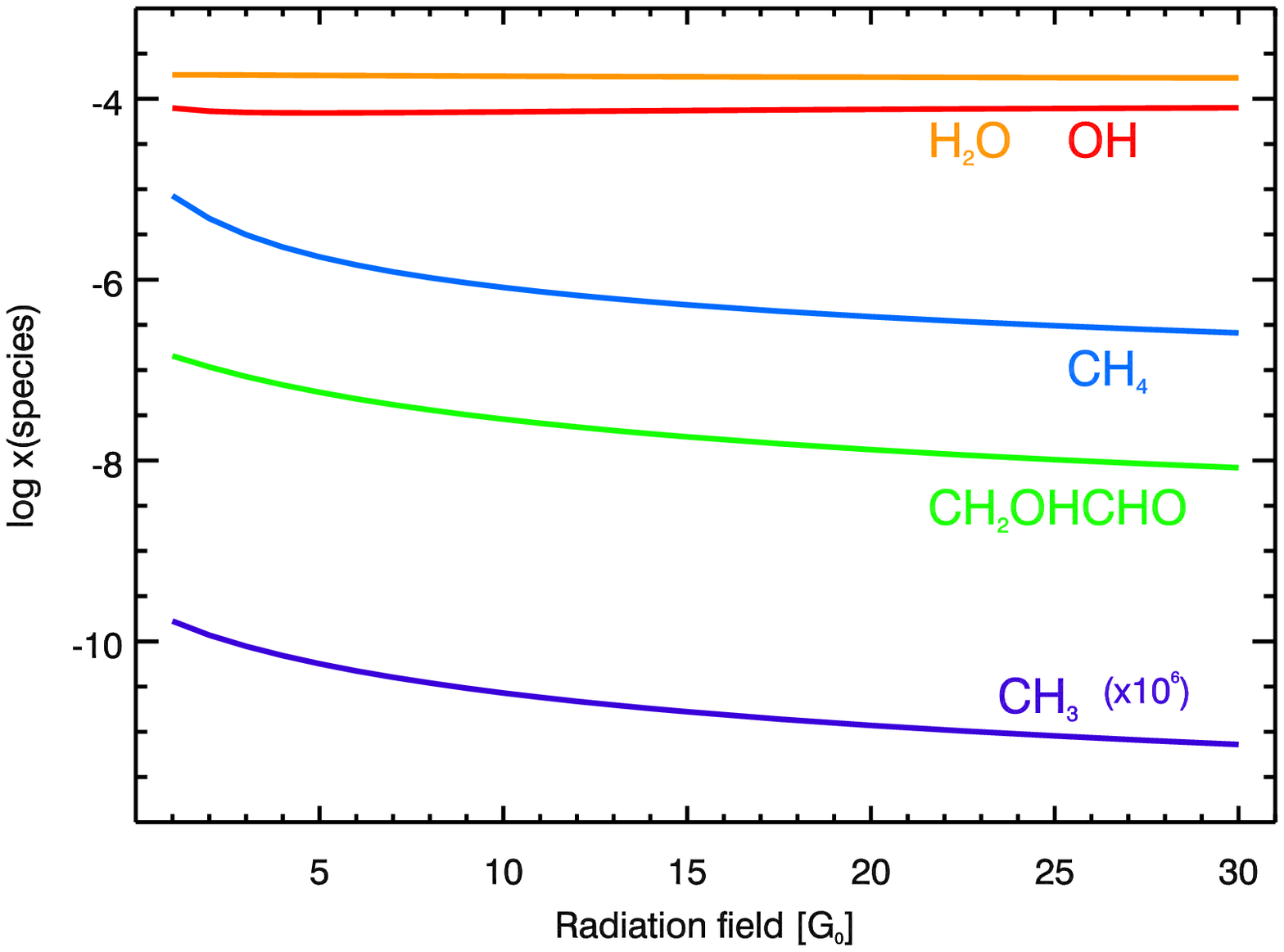}
\caption{The dependance of glycolaldehyde abundance upon UV radiation
field intensity, for mechanism A. See the electronic edition
  of the Journal for a color version of this figure. \label{fig:mechAUV}}
\end{figure}

In order to explore the effects of enhanced UV irradiation of surface
ices in mechanism A (Table~\ref{tab:schemes}), we have used the
standard reaction rates as shown in Table~\ref{tab:rates}, and the
more conservative hydrogenation regime (Table~\ref{tab:hydrog}). We
varied the strength of the UV field by up to a factor of 30 over the
standard interstellar field, and the results are plotted in
Fig.~\ref{fig:mechAUV}. Intuitively one would expect that increased
grain processing of surface-bound H$_2$O and CH$_4$\ would lead to a
greater production of glycolaldehyde. However, higher UV fluxes mean
that the gas-phase species which go on to form CH$_4$\ in particular
on the grains are destroyed by reaction with abundant
photodissociation products like H$_3^+$\ and H$^+$. Thus the limited
abundance of grain-surface CH$_4$\ limits the formation of
glycolaldehyde. Water ice increases very marginally in abundance when
the core is under higher levels of irradiation.

\subsection{Reaction rate analysis of mechanism A}

\begin{figure*}
\epsscale{1.795}
\plotone{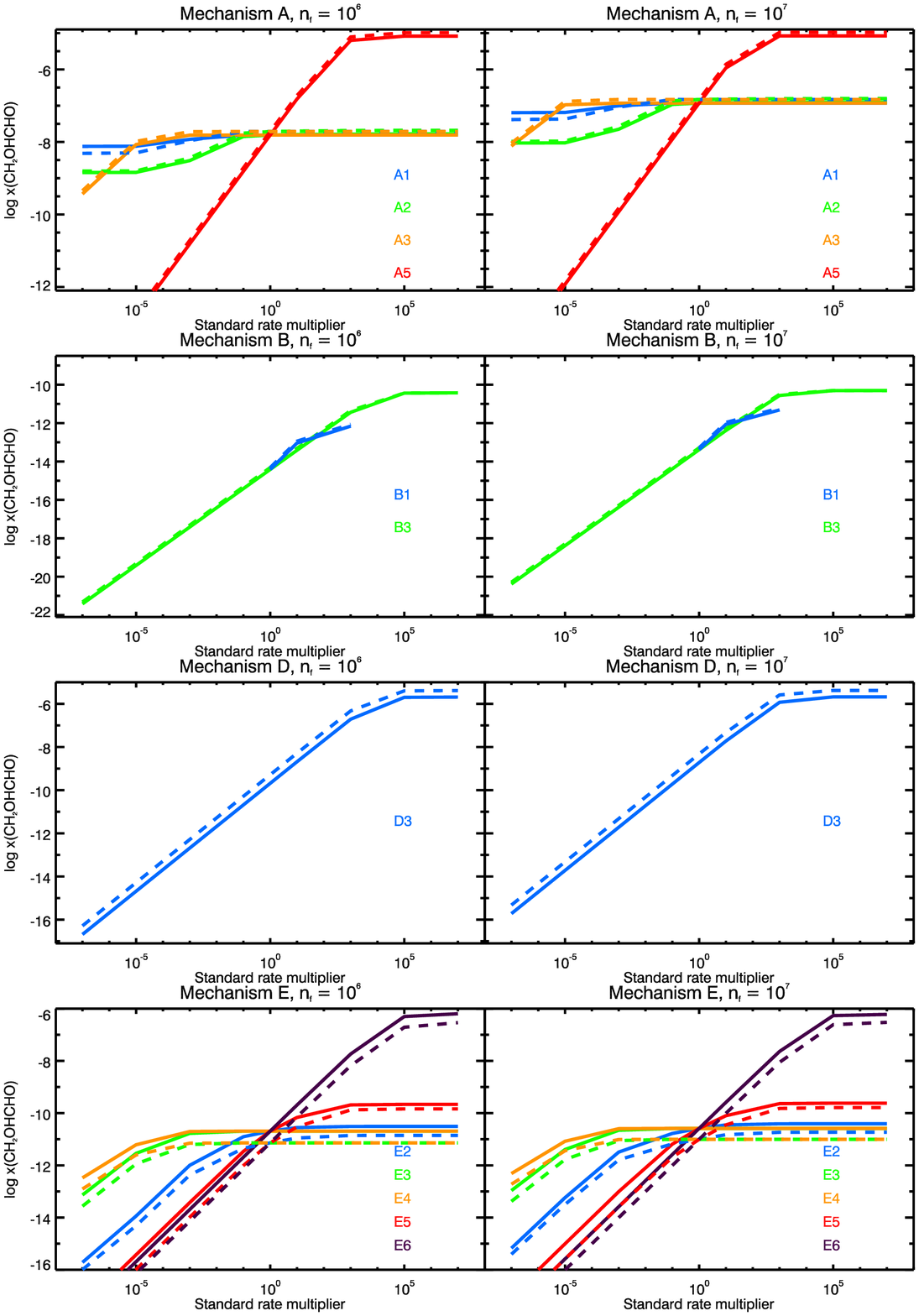}
\caption{The production of glycolaldehyde via mechanisms A, B, D \& E,
  for $n_\mathrm{f}$=10$^6$\ and 10$^7$\,cm$^{-3}$\ (see
  \S\ref{sec:mechC} for a description of mechanism C). The solid
  lines show results using hydrogenation regime f1; the dashed lines,
  regime f2. Standard rates can be found in Table~\ref{tab:rates}. See
  the electronic edition of the Journal for a color version of this
  figure. \label{fig:mechall}}
\end{figure*}

Results of the model can be found in Fig.~\ref{fig:mechall}, where we
plot the fractional abundance of glycolaldehyde obtained at the final
density of the collapse, $n_\mathrm{f}$, against the scale factor of
the \textquotedblleft standard\textquotedblright\ rates, as found in
Table~\ref{tab:rates}. We compare models with differing
$n_\mathrm{f}$\ and in the two different hydrogenation regimes, f1 and
f2. The fractional abundance of glycolaldehyde in the model is highly
dependent on the rate of the final reaction of mechanism A, A5 -- it
scales linearly with the rate until 1--10 times the standard rate,
after which the curve turns over to a maximum abundance of
$x$(\gal)$\sim$10$^{-5}$. The effect of the differing hydrogenation
regimes is minimal. The fractional abundance of glycolaldehyde is
fairly insensitive to the rates of reactions A1--A3, increasing by
only a factor of 10 or so whilst the rate coefficients span a range of
10$^{14}$. Reactions A1 and A2 compete with grain surface
hydrogenation of O and C in the provision of OH and CH$_3$\ radicals,
respectively. As core density increases, photons become increasingly
absorbed, meaning that reaction A3 proceeds with reactants that are
products of grain-surface hydrogenation rather than grain-surface
photolysis \citep[cf.,][]{pee06}. Reaction A3 itself is competing with
the hydrogenation of adsorbed CO, which dominates at high densities,
meaning that the abundance of glycolaldehyde is relatively independent
of reactions A1--A3 in general.

Reaction A3 has been studied in the literature, as part of
investigations into methanol formation. Experiments on H$_2$O-CO ice
at $T<20$\,K by \citet{hid04} do not show any evidence of methane
production, which implies that abundances of CH$_3$\ are low. It seems
likely that methanol production results mainly from the successive
hydrogenation of CO (i.e., CO $\longrightarrow$\ HCO
$\longrightarrow$\ H$_2$CO $\longrightarrow$\ CH$_3$O
$\longrightarrow$\ CH$_3$OH) rather than via reaction A3
\citep{hid04}.

Reaction A4 has also been well-studied in the literature
\citep[e.g.,][]{hud99,wat03}, and is a crucial reaction in many of the
mechanisms studied here (it is identical to B2, D2 and E1). HCO ice
has not been detected in the interstellar medium, implying that its
formation is slower than subsequent reactions, e.g., $\mathrm{H} +
\mathrm{HCO} \longrightarrow\ \mathrm{H_2CO}$, which have lower
activation energies \citep{wat02}. Indeed, the formation of H$_2$CO in
this way has been shown to be barrierless \citep{gou07}. \mbox{CO + H}
has a barrier of several thousand Kelvin in the gas phase \citep[see
  summary by][]{hid07}, but barriers on a surface depend on the
composition and structure of that surface \citep{wat04,gou08}, with in
some cases the reaction being completely barrierless
\citep{gou08}. \citet{hid07} calculate a rate for A4 from an
experiment involving various combinations of CO and H$_2$O ice. They
find that the product k$_\mathrm{H}$n$_\mathrm{H}$\ is
$\sim$5$\times$10$^{-3}$\,s$^{-1}$, which for 10$^{-3}$\ H atoms per
square centimetre of surface (n$_\mathrm{H}$; typical for a large
grain) gives a reaction timescale on the order of seconds. For small
grains, the timescale could be on the order of a year (i.e., fast by
astrophysical standards).

Reaction A5 requires the diffusion of molecules and radicals across a
grain surface, which is a slow process at 10\,K. However, experimental
results show that there is some evidence of complex surface reactions,
even at 10\,K \citep[e.g.,][]{wat02}. Figure~\ref{fig:mechall}
suggests that even at rates slower than our conservative standard
rate, significant amounts of glycolaldehyde form via this mechanism.

Increasing $n_\mathrm{f}$\ by an order of magnitude from 10$^6$\ to
10$^7$\,cm$^{-3}$\ has the effect of increasing glycolaldehyde
production for a given reaction rate by a corresponding order of
magnitude, approximately. This reflects the greater collisional rate
between molecules and grains, and thus a greater freeze-out rate. The
peak fractional abundance of glycolaldehyde produced via mechanism A,
$x$(\gal)$\sim$10$^{-5}$, is unaffected by the change in
$n_\mathrm{f}$, showing that a significant proportion of the available
carbon ends up in glycolaldehyde at the most extreme rates
investigated, something which is unlikely to occur naturally.

\subsection{Reaction rate analysis of mechanism B}

The reaction mechanism suggested by \citet{ben07b}, which was
identified experimentally in the laboratory, is very inefficient at
producing glycolaldehyde at 10\,K. Even under conditions where the
cosmic ray ionisation rate is increased by seven orders of magnitude,
less than $x$(\gal)$\sim$10$^{-9}$\ results
(Fig.~\ref{fig:mechall}). The rates adopted for reactions B1 and B3
are critical to the amount of glycolaldehyde produced, with the yield
scaling linearly with the adopted rate. Using our standard rates for
B1 and B3, $x$(\gal)$\sim$10$^{-13}$--10$^{-14}$, significantly lower
than that observed in \gto\ \citep{bel09}. Moreover, we do not include
the hydrogenation of the hydroxymethyl (CH$_2$OH) radical to methanol
in our reaction scheme, which surely must be rapid and compete with
reaction B3.

\subsection{Reaction rate analysis of mechanism C}
\label{sec:mechC}

Mechanism C, the only gas-phase reaction mechanism we investigate,
also is not particularly efficient in the production of
glycolaldehyde, producing $x$(\gal)$\lesssim$10$^{-10}$ at the most
enhanced values of the reaction rate coefficients. We do not increase
the rate coefficient of reaction C1 beyond
6.3$\times$10$^{-7}$\,cm$^3$\,s$^{-1}$, since this would be incredibly
fast for a gas-phase reaction. In fact, we only vary the rate of C1
for completeness, since the standard reaction rate
\citep[from][]{tan79} is accurate to 25\%\ according to the UDfA
database. Both reactions C1 and C2 produce linearly increasing amounts
of \gal\ with increasing reaction rate coefficient, producing
$x$(\gal)$=$10$^{-13}$ at the standard rates. The limiting factor in
this mechanism is the availability of gas-phase formaldehyde, which at
10\,K is only 1\% of the total formaldehyde, the rest being frozen
onto grain surfaces. However, we do not include non-thermal desorption
mechanisms in our simple model, which could increase the amount of
formaldehyde in the gas phase. \citet{rob07} show that non-thermal
desorption can return a large proportion of H$_2$CO to the gas phase
in extreme cases. It is not clear how their results would apply to our
situation, where the density is larger, and thus freezeout more rapid.

\subsection{Reaction rate analysis of mechanism D}

This grain-surface reaction based on the work of \citet{bel09}
involves the products of rapid hydrogenation of CO and HCO$^{(+)}$,
meaning that the only instructive reaction rate to investigate is that
of reaction D3. We simplify the three-body reaction proposed by
\citet{bel09} with a two-body reaction by assuming that the H-atom
addition is rapid. A maximum fractional abundance of glycolaldehyde of
$\sim$10$^{-6}$\ is produced when the standard rate for the reaction
is increased by 1\,000--10\,000, depending on $n_\mathrm{f}$. At the
standard rate, $x$(\gal)$\approx$10$^{-9\ldots-10}$, somewhat smaller
the derived from their observations, but within their error
constraints (see Sect.~\ref{sec:discuss}).

\subsection{Reaction rate analysis of mechanism E}

The atom-addition mechanism suggested by \citet{cha05} is the most
complex that we have considered, involving six two-body
reactions. Three of these reactions are reactions with H, which are
assumed rapid, but the surface migration of heavier atoms is
significantly slower due to the large diffusion barriers involved
\citep[e.g.,][]{lei84}. The rates of reactions E2--E4 have little
effect on the abundance of glycolaldehyde, showing largely flat
profiles in Fig.~\ref{fig:mechall}. The gradient of reaction E5, where
an oxygen atom is added to the molecule, is somewhat steeper, but
reaction E6 is the crucial reaction in this mechanism. We have assumed
a fairly conservative value of 3$\times$10$^{-17}$\,s$^{-1}$\ for this
reaction, due to a potential barrier in the H-addition process, but if
the reaction were 10$^{3\ldots5}$\ times faster than expected, it
could produce $x$(\gal)$\sim$10$^{-6}$.

\subsection{Further experimentation}

\begin{figure}
\epsscale{1.0}
\plotone{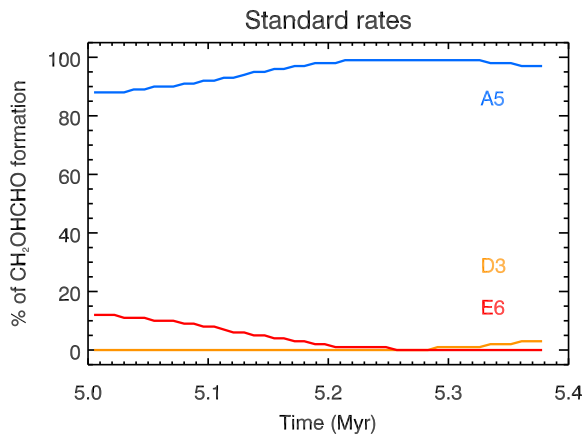}
\caption{The formation route of glycolaldehyde with all mechanisms in
  operation, using standard reaction rates. See the electronic edition
  of the Journal for a color version of this figure. \label{fig:max}}
\end{figure}

The analysis performed thus far has been somewhat artificial, since if
the reactions contained in the suggested mechanisms occur, then they
likely occur in competition with each other (and other reactions),
i.e., reactions in other mechanisms are not \textquoteleft switched
off\textquoteright. To investigate this, we performed two further
experiments where we set the rates of all the reactions considered to
their standard rates, and where we set them to \textquotedblleft
optimal\textquotedblright\ rates. By optimal here we mean adopting
$\alpha$ values for where the rate profiles in Fig.~\ref{fig:mechall}
turn over to become flat. Rates for mechanism C were kept as standard,
since the rate profiles do not turn over.

We find that when using standard rates for all reactions,
$x$(\gal)$=1.5\times10^{-7}$. In this competitive environment,
reactions A5 and E6 are the dominant routes to the formation of
glycolaldehyde, with A5 being the most efficient by a considerable
margin as the core collapses (see Fig.~\ref{fig:max}, left). Using
\textquotedblleft optimal\textquotedblright\ rates an extremely large
amount of glycolaldehyde can form, $x$(\gal)$=1.1\times10^{-5}$, which
is approximately 6\%\ of the elemental carbon abundance of the
core. In this case the most dominant reaction for its formation at
later times is B3 (the rate of which has been enhanced 10$^5$\ times);
E6 dominates at early times, with a rate enhancement of 10$^7$. Given
these extreme rate enhancements, this scenario is unlikely.

\section{Discussion}
\label{sec:discuss}

Having investigated the formation of glycolaldehyde via five different
reaction mechanisms, it is clear that considerable quantities of
glycolaldehyde can be produced. Also, very small amounts of
glycolaldehyde can result from mechanisms which are inefficient at
10\,K (e.g., mechanism B). Many of the rates involved in these
mechanisms are completely unknown, and others have a large degree of
uncertainty. Given that fractional abundances of glycolaldehyde can
vary over ten orders of magnitude or more in the models, further work
needs to be done in order to fully understand how glycolaldehyde forms
at low temperatures in dense molecular cores, like those in \gto. We
are currently undertaking Density Function Theory calculations for
these five mechanisms, with the results to be forthcoming in a future
publication.

The fact that glycolaldehyde has been detected in \gto\ provides us
with some constraints on our modelling. \citet{bel09} estimated that
the fractional abundance of glycolaldehyde towards this region was on
the order of 10$^{-8}$. Due to the uncertainties of temperature and
column density in their measurements, the errors in this estimate are
large: potentially two orders of magnitude \citetext{M. Beltr\'an,
  priv. comm.}. This lower limit of
$x$(\gal)$\gtrsim$10$^{-10}$\ effectively means that we can reasonably
exclude mechanisms B and C from further consideration, since they do
not produce enough glycolaldehyde even under the most favourable
conditions. Mechanism E only produces required abundances if the rate
of the final reaction in the scheme is enhanced.  Mechanisms A and D
have greater fecundity, and Fig.~\ref{fig:max} shows that mechanism A
is considerably more efficient when in competition. Thus it appears
that for the rates we have adopted, the main formation mechanism for
glycolaldehyde at 10\,K is:
\begin{eqnarray}
\nonumber \mathrm{CH_3} + \mathrm{OH} &\longrightarrow& \mathrm{CH_3OH}\\
\nonumber \mathrm{CO} + \mathrm{H} &\longrightarrow& \mathrm{HCO}\\
\nonumber \mathrm{CH_3OH} + \mathrm{HCO} &\longrightarrow& \mathrm{CH_2OHCHO} + \mathrm{H}
\end{eqnarray}
on grain surfaces. The formation of both CH$_3$OH and HCO in ices has
been well-studied experimentally
\citep[e.g.,][]{wat04,hid04,wat03,wat02}, and is related, with HCO
being a crucial part in the formation of CH$_3$OH, which is the
terminal molecule in the hydrogenation process. Only energetic
processes (e.g., UV or cosmic-ray irradiation) will decompose
CH$_3$OH, since H-abstraction to form H$_2$CO is negligible
\citep{hid04}.

\subsection{A chemical evaluation of mechanisms A, D and E}
\label{sec:dft}

Building on the astrophysical models, we now consider reactions A, D
and E from a physicochemical perspective, taking account of the
intrinsic thermodynamic stability of reagents and products. A full
treatment of this is underway, but due to the
computationally-expensive nature of such an investigation, here we
only consider general principles. Firstly, in mechanism A, reaction A5
is found to be particularly efficient at producing glycolaldehyde:
\begin{eqnarray}
\nonumber \rm{CH_3OH} + \rm{HCO} \longrightarrow\ \rm{CH_2OHCHO}\ + \rm{H}.
\end{eqnarray}
However, A5 is the reaction of a stable molecule (methanol) with a
reactive radical (formyl) to give stable glycolaldehyde and an H
monoatom. The H monoatom is extremely reactive and even at low
temperatures is very mobile, hence the rate of the backward reaction
(addition of H to glycolaldehyde) could be expected to be competitive
with that of the forward reaction. Preliminary {\it ab-initio}
calculations conducted at the coupled-cluster level with single,
double and triple excitations (CCSD(T)) and a triple-zeta quality
basis set shows that in fact reaction A5 is endothermic, and
consequently the rate of the reaction yielding glycolaldehyde would be
very unfavourable.

Reaction D3, 
\begin{eqnarray}
\nonumber  \rm{H_2CO} + \rm{HCO} + \rm{H} \longrightarrow\ \rm{CH_2OHCHO},
\end{eqnarray}
is a three-body reaction which has a vanishingly small probability of
occurring, further hindered by the low temperatures considered
here. However, the products could be obtained by two sequential
two-step reactions: first, reaction of H with HCO yields H$_2$CO, a
barrierless process in the gas phase, according to published
theoretical work \citep{gou07}, which could combine with another
H$_2$CO molecule to yield the glycolaldehyde product. However, H$_2$CO
is rather stable and hence the reaction rate for the condensation of
two H$_2$CO molecules could be expected to be rather slow and hence
improbable.  A second possibility is the reaction of H with H$_2$CO
which according to past work \citep{woo02, sae83} yields H$_3$CO as
the kinetic product. H$_3$CO could react with HCO to give
glycolaldehyde, and this reaction ought to have a low barrier since
both H$_3$CO and HCO are reactive radical species.

In mechanism E, the final promising path identified from the
astrophysical models, glycoladehyde is assembled from a building block
of CO via six stepwise monatomic addition reactions. Although the
constituent reactions of mechanism E are chemically viable,
consideration of the physical conditions and the reaction
probabilities suggest that this pathway may be an unlikely source of
glycolaldehyde. Under the conditions of a temperature of 10\,K, only
monatomic H is mobile; C and O are static and only become mobile
during warm-up, which implies that reactions E2 and E5 are highly
unlikely to occur. Hence these reactions are likely to be
rate-limiting. Construction of glycolaldehyde via monoatomic addition
also depends on a well-defined consecutive set of reactions. In this
scheme, HCO reacts with monatomic C (reaction E2), yet monatomic H is
present at higher abundance and is known to react without a barrier
\citep{gou07} with HCO, the product of E1, to give H$_2$CO rather the
product of E2, HC$_2$O.  Since mechanism E is composed of six
reactions which are expected to be limited by at least two of those
reactions (E2 and E5), we consider this pathway not to be very
probable or efficient.

Of the reactions identified by the astrophysical models, it is
suggested that mechanism D is most probable according to chemical
considerations.  Detailed ab-initio calculations are underway to
assess the influence of substrates on the reaction barriers and hence
quantify the efficiency of mechanisms A, D and E. It should be noted
that mechanisms B and C are viable from a chemical standpoint and
these will also be considered in comparison to A, D and E to assess
the most viable scheme.

\section{Summary}
\label{sec:concs}

We have investigated five reaction pathways for the formation of
glycolaldehyde, a simple sugar, which have been previously suggested
in the astrophysical literature, but not thoroughly tested or
justified. By means of a chemical model of an isothermally-collapsing
molecular core, we have determined that under the physical conditions
assumed, mechanisms B and C \citep[suggested
  by][respectively]{ben07b,hal06} are relatively inefficient and are
unlikely to be the major pathways to the formation of
glycolaldehyde. Mechanisms D and E \citep[suggested
  by][respectively]{bel09,cha05} can reach observed fractional
abundances of glycolaldehyde if reaction rates are enhanced by factors
of 100 or more over those we have chosen as standard. Finally,
mechanism A, from the work of \citet{sor01}, can produce
glycolaldehyde very efficiently; however, in the high density regions
around forming protostars, photons are unlikely to penetrate enough to
initiate the reaction scheme as it is. Instead, the initial reactants,
OH and CH$_3$, are amply supplied by the freeze-out and subsequent
hydrogenation of atoms and smaller radicals.

Further evaluation of the reaction schemes taking reagent and product
stability into account leads us to expect that mechanism A may not be
as efficient as the astrophysical modelling suggests because of a high
barrier for the final step in the formation of glycoladehyde. However,
mechanism D may be more likely, provided the final reaction in the
scheme proceeds as two two-body reactions involving H$_3$CO and HCO
radicals, rather than a single three-body reaction. Detailed
examination of the reaction barriers according to high level quantum
chemical methods is currently being undertaken, taking into account
the role of different substrates.

Finally, the list of reactions considered here is not exhaustive and
we are seeking to use astrophysical modelling, quantum chemical and
experimental approaches to identify whether other, as yet unreported,
pathways are more efficient at producing glycolaldehyde.

\acknowledgments

The authors appreciate fruitful discussions with T.~P.~M. Goumans,
which have benefitted the content of this paper. Funding for this work
was provided by the Leverhulme Trust to PMW and DJB. FP acknowledges
support from the LASSIE Initial Training Network under the European
Community's Seventh Framework Programme FP7/2007-2013 under grant
agreement \#238258.

\end{document}